\documentclass[12pt]{article}
  \usepackage{amsfonts}
  \usepackage{amsmath}
\usepackage{amssymb}
\usepackage{amscd}
\usepackage{graphicx}
\begin{document}

~~
\bigskip
\bigskip
\begin{center}
{\Large {\bf{{{Two-particle system in Coulomb potential for twist-deformed space-time}}}}}
\end{center}
\bigskip
\bigskip
\bigskip
\begin{center}
{{\large ${\rm {Marcin\;Daszkiewicz}}$}}
\end{center}
\bigskip
\begin{center}
\bigskip

{ ${\rm{Institute\; of\; Theoretical\; Physics}}$}

{ ${\rm{ University\; of\; Wroclaw\; pl.\; Maxa\; Borna\; 9,\;
50-206\; Wroclaw,\; Poland}}$}

{ ${\rm{ e-mail:\; marcin@ift.uni.wroc.pl}}$}

\end{center}
\bigskip
\bigskip
\bigskip
\bigskip
\bigskip
\bigskip
\bigskip
\bigskip
\bigskip
\begin{abstract}
In this article, we define two-particle system in Coulomb potential for twist-deformed space-time with spatial directions commuting to time-dependent function $f_{\kappa_a}({t})$. Particularly, we provide the proper Hamiltonian function and further, we rewrite it in terms of commutative variables. Besides, we demonstrate, that for small values of deformation parameters, the obtained in the perturbation framework, first-ordered corrections for ground Helium energy are equal to zero. In such a way we show, that nontrivial impact of space-time noncommutativity appears only at the second order of the quantum-mechanical correction expansion.
\end{abstract}
\bigskip
\bigskip
\bigskip
\bigskip
\eject

\section{Introduction}

The suggestion to use noncommutative coordinates goes back to
Heisenberg and was firstly  formalized by Snyder in \cite{snyder}.
Recently, there were also found formal  arguments based mainly  on
Quantum Gravity \cite{2}-\cite{2a} and String Theory models
\cite{recent}-\cite{string1}, indicating that space-time at the Planck
scale  should be noncommutative, i.e., it should  have a quantum
nature. Consequently, there appeared a lot of papers dealing with
noncommutative classical and quantum  mechanics (see e.g.
\cite{mech}-\cite{qm}) as well as with field theoretical models
(see e.g. \cite{prefield}-\cite{field}), in which  the quantum
space-time is employed.

In accordance with the Hopf-algebraic classification of all
deformations of relativistic \cite{class1} and nonrelativistic
\cite{class2} symmetries, one can distinguish three basic types
of space-time noncommutativity (see also \cite{nnh} for details):\\
\\
{ \bf 1)} Canonical ($\theta^{\mu\nu}$-deformed) type of quantum space \cite{oeckl}-\cite{dasz1}
\begin{equation}
[\;{ x}_{\mu},{ x}_{\nu}\;] = i\theta_{\mu\nu}\;, \label{noncomm}
\end{equation}
\\
{ \bf 2)} Lie-algebraic modification of classical space-time \cite{dasz1}-\cite{lie1}
\begin{equation}
[\;{ x}_{\mu},{ x}_{\nu}\;] = i\theta_{\mu\nu}^{\rho}{ x}_{\rho}\;,
\label{noncomm1}
\end{equation}
and\\
\\
{ \bf 3)} Quadratic deformation of Minkowski and Galilei  spaces \cite{dasz1}, \cite{lie1}-\cite{paolo}
\begin{equation}
[\;{ x}_{\mu},{ x}_{\nu}\;] = i\theta_{\mu\nu}^{\rho\tau}{
x}_{\rho}{ x}_{\tau}\;, \label{noncomm2}
\end{equation}
with coefficients $\theta_{\mu\nu}$, $\theta_{\mu\nu}^{\rho}$ and  $\theta_{\mu\nu}^{\rho\tau}$ being constants.\\
\\
Moreover, it has been demonstrated in \cite{nnh}, that in the case of the
so-called N-enlarged Newton-Hooke Hopf algebras
$\,{\mathcal U}^{(N)}_0({ NH}_{\pm})$ the twist deformation
provides the new  space-time noncommutativity of the
form\footnote{$x_0 = ct$.},\footnote{The discussed space-times have been  defined as the quantum
representation spaces, so-called Hopf modules (see e.g. \cite{oeckl}, \cite{chi}), for the quantum N-enlarged
Newton-Hooke Hopf algebras.},\footnote{The indices $+$ and $-$ in functions $f_{+}\left(\frac{t}{\tau}\right)$ and
$f_{-}\left(\frac{t}{\tau}\right)$ correspond to N-enlarged Newton-Hooke Hopf algebras
$\,{\mathcal U}^{(N)}_0({ NH}_{+})$ and $\,{\mathcal U}^{(N)}_0({ NH}_{-})$ respectively.}
\begin{equation}
{ \bf 4)}\;\;\;\;\;\;\;\;\;[\;t,{ x}_{i}\;] = 0\;\;\;,\;\;\; [\;{ x}_{i},{ x}_{j}\;] = 
if_{\pm}\left[\frac{t}{\tau}\right]\theta_{ij}(x)
\;, \label{nhspace}
\end{equation}
with time-dependent  functions
$$f_+\left[\frac{t}{\tau}\right] =
f\left[\sinh\left[\frac{t}{\tau}\right],\cosh\left[\frac{t}{\tau}\right]\right]\;\;\;,\;\;\;
f_-\left[\frac{t}{\tau}\right] =
f\left[\sin\left[\frac{t}{\tau}\right],\cos\left[\frac{t}{\tau}\right]\right]\;,$$
$\theta_{ij}(x) \sim \theta_{ij} = {\rm const}$ or
$\theta_{ij}(x) \sim \theta_{ij}^{k}x_k$ and  $\tau$ denoting the time scale parameter
 -  the cosmological constant. Besides, it should be  noted, that the  above mentioned quantum spaces {\bf 1)}, { \bf 2)} and { \bf 3)}
can be obtained  by the proper contraction limit  of the commutation relations { \bf 4)}\footnote{Such a result indicates that the twisted N-enlarged Newton-Hooke Hopf algebra plays a role of the most general type of quantum group deformation at nonrelativistic level.}.

In this article, we define two-particle system in Coulomb potential for twist-deformed space-time \cite{nnh} with $\theta_{ij}(x) \sim \theta_{ij}$\footnote{For such a construction in the case of most simple (canonical) type of space-time noncommutativity see \cite{stare}.}.
Particularly, we provide the proper Hamiltonian function and further, we rewrite it in terms of commutative variables. Besides, we demonstrate, that for small values of deformation parameters, the obtained in the perturbation framework, first-ordered corrections for ground Helium energy are equal to zero. In such a way we show, that nontrivial impact of space-time noncommutativity appears only at the second order of the quantum-mechanical correction expansion.

The paper is organized as follows. In second Section we remaind the basic facts concerning the twisted, nonrelativistic spaces \cite{nnh}. In the next Section, we recall the perturbation calculations of ground energy for Helium atom in the case of classical space-time. In the fourth Section, which is devoted to their twisted counterpart, we demonstrate, that first-ordered corrections vanish. The discussion and final remarks are presented in the last Section.

\section{Twist-deformed space-times}

In this Section we turn closer to the twisted N-enlarged Newton-Hooke spaces equipped with two spatial directions
commuting to classical time, i.e., we consider spaces of the form \cite{nnh}
\begin{equation}
[\;t,\hat{x}_{i}\;] =[\;\hat{x}_{1},\hat{x}_{3}\;] = [\;\hat{x}_{2},\hat{x}_{3}\;] =0\;\;\;,\;\;\;[\;\hat{x}_{1},\hat{x}_{2}\;] = if({t})\;\;;\;\;i=1,2,3\;, \label{spaces}
\end{equation}
with function $f({t})$ given for example in the most simple $N=1$ case by
\begin{eqnarray}
f({t})&=&f_{\kappa_1}({t}) =
f_{\pm,\kappa_1}\left[\frac{t}{\tau}\right] = \kappa_1\,C_{\pm}^2
\left[\frac{t}{\tau}\right]\;, \label{w2}\\
f({t})&=&f_{\kappa_2}({t}) =
f_{\pm,\kappa_2}\left[\frac{t}{\tau}\right] =\kappa_2\tau\, C_{\pm}
\left[\frac{t}{\tau}\right]S_{\pm} \left[\frac{t}{\tau}\right] \;,
\label{w3}\\
f({t})&=&f_{\kappa_3}({t}) =
f_{\pm,\kappa_3}\left[\frac{t}{\tau}\right] =\kappa_3\tau^2\,
S_{\pm}^2 \left[\frac{t}{\tau}\right] \;, \label{w4}\\
f({t})&=&f_{\kappa_4}({t}) =
 f_{\pm,\kappa_4}\left[\frac{t}{\tau}\right] = 4\kappa_4
 \tau^4\left[C_{\pm}\left[\frac{t}{\tau}\right]
-1\right]^2 \;, \label{w5}\\
f({t})&=&f_{\kappa_5}({t}) =
f_{\pm,\kappa_5}\left[\frac{t}{\tau}\right] = \pm \kappa_5\tau^2
\left[C_{\pm}\left[\frac{t}{\tau}\right]
-1\right]C_{\pm} \left[\frac{t}{\tau}\right]\;, \label{w6}
\end{eqnarray}
\begin{eqnarray}
f({t})&=&f_{\kappa_6}({t}) =
f_{\pm,\kappa_6}\left[\frac{t}{\tau}\right] = \pm \kappa_6\tau^3
\left[C_{\pm}\left[\frac{t}{\tau}\right] -1\right]S_{\pm}
\left[\frac{t}{\tau}\right]\;; \label{w7}
\end{eqnarray}
\begin{eqnarray}
C_{+} \left[\frac{t}{\tau}\right] = \cosh\left[\frac{t}{\tau}\right]\;\;\;&{\rm and}&\;\;\;
S_{+} \left[\frac{t}{\tau}\right] = \sinh
\left[\frac{t}{\tau}\right] \;,\nonumber\\
C_{-} \left[\frac{t}{\tau}\right] = \cos \left[\frac{t}{\tau}\right]\;\;\;&{\rm and}&\;\;\;
S_{-} \left[\frac{t}{\tau}\right] = \sin
\left[\frac{t}{\tau}\right] \;.\nonumber
\end{eqnarray}
As it was already mentioned in Introduction, in $\tau \to \infty$ limit the above quantum spaces reproduce the canonical (\ref{noncomm}),
Lie-algebraic (\ref{noncomm1}) as well as quadratic (\ref{noncomm2})  type of
space-time noncommutativity, with
\begin{eqnarray}
f_{\kappa_1}({t}) &=& \kappa_1\;,\label{nw2}\\
f_{\kappa_2}({t}) &=& \kappa_2\,t\;,\label{nw3}\\
f_{\kappa_3}({t}) &=& \kappa_2\,t^3\;,\label{nw4}\\
f_{\kappa_4}({t}) &=& \kappa_4\,t^4\;,  \label{nw5}\\
f_{\kappa_5}({t}) &=& \frac{1}{2}\kappa_5\,t^2\;, \label{nw6}\\
f_{\kappa_6}({t}) &=& \frac{1}{2}\kappa_6\,t^3\;. \label{nw7}
\end{eqnarray}
Moreover, let us notice that the spaces (\ref{spaces}) can be extended to the case of multiparticle systems as follows
\begin{eqnarray}
&&[\;t,\hat{x}_{iA}\;] =[\;\hat{x}_{1A},\hat{x}_{3B}\;] = [\;\hat{x}_{2A},\hat{x}_{3B}\;] =
0\;\;;\;\;i=1,2,3\;,\label{spaces300}\\
&&\nonumber\\
&&~~~~~~~~~~~~~[\;\hat{x}_{1A},\hat{x}_{2B}\;] =
if_{\kappa_a}({t})\delta_{AB}\;,  \nonumber
\end{eqnarray}
with $A, B = 1,2, \ldots ,M$. Besides, it should be observed that  such an extension (blind in $A$, $B$ indecies) is compatible with canonical deformation (\ref{noncomm}).
Precisely, in $\tau$ approaching infinity limit the space (\ref{spaces300}) with function $f_{\kappa_a}(t) = f_{\pm,\kappa_1}\left(\frac{t}{\tau}\right) = \kappa_1\,C_{\pm}^2
\left(\frac{t}{\tau}\right)$ passes into the well-known multiparticle canonical space-time proposed in \cite{fiore}\footnote{It should be noted that  modification of the relation (\ref{spacesfiore}) (blind in $A$, $B$ indieces as well) is in accordance with the formal arguments proposed in \cite{fiore}. Precisely, the relations (\ref{spacesfiore}) are constructed with adopt so-called braided tensor algebra procedure, dictated by structure of quantum $R$-matrix for canonical deformation \cite{qg1}, \cite{qg3}. We would like to mention, however, that in \cite{fiore} an
erroneous conclusion has been stated that based on such a twisted symmetry the noncommutative
quantum field theory (QFT) on the quantum space satisfying the relation in (\ref{noncomm}), and the usual commutative QFT are
identical. This conclusion has been reached by a misuse of the proper transformation properties of the
fields in the corresponding noncommutative space time \cite{chaia}.} (see also \cite{odynswiatowid})
\begin{equation}
[\;t,\hat{x}_{iA}\;] =[\;\hat{x}_{1A},\hat{x}_{3B}\;] = [\;\hat{x}_{2A},\hat{x}_{3B}\;] =
0\;\;\;,\;\;\; [\;\hat{x}_{1A},\hat{x}_{2B}\;] =
i\kappa_1\delta_{AB}
\;. \label{spacesfiore}
\end{equation}
Obviously, for $f(t)$ approaching zero all above spaces become classical.

\section{Two-particle $(M=2)$ Coulomb system in commutative space-time}

In this Chapter we recall the facts concerning the model of two particles mutually interacting and moving in the presence of Coulomb potential. Then, in accordance with \cite{lit1}-\cite{qwerty}, the corresponding Hamiltonian function is given by
\begin{eqnarray}
H\left[\bar{p}_1,\bar{p}_2;\bar{r}_1,\bar{r}_2\right] = H_0 \left[\bar{p}_1,\bar{p}_2;\bar{r}_1,\bar{r}_2\right]+H_{\rm int}\left[\bar{r}_1,\bar{r}_2\right]\;,\label{ham}
\end{eqnarray}
where
\begin{eqnarray}
H_0 \left[\bar{p}_1,\bar{p}_2;\bar{r}_1,\bar{r}_2\right]=\frac{1}{2}\left[\;{\bar{p}_1^2}/{m_1}+{\bar{p}_2^2}/{m_2}\;\right]-\frac{Qq_1}{r_1}-
\frac{Qq_2}{r_2}\;,\label{ham1}
\end{eqnarray}
and
\begin{eqnarray}
H_{\rm int}\left[\bar{r}_1,\bar{r}_2\right]=\frac{q_1q_2}{r_{12}}\;\;\;;\;\;\;r_{12} = |\bar{r}_1-\bar{r}_2|\;,\label{ham2}
\end{eqnarray}
with $m_1$, $m_2$, $q_1$, $q_2$ and $Q$ denoting the masses and charges of the particles and the charge of the source respectively. It should be also noted, that the first, $H_0$-term in the formula (\ref{ham}) describes the sum of two energy operators for hydrogen atoms, while the second one - the interaction of the particles called electrostatic repulsion.

It is well-known, that if one neglects $H_{\rm int}$-term then the eigenvectors and eigenvalues of $H$-operator
\begin{eqnarray}
H\psi = H_0\psi = E\psi\;, \label{dixiland}
\end{eqnarray}
are given by\footnote{We put $Q=Ze,\;q_1=q_2=e$ as well as $m_1=m_2=m_{\rm e}$ where $e$ and $m_{\rm e}$ are the charge and mass of the electron.}
\begin{eqnarray}
\psi&=&\psi_{(n_1,l_1,m_1)}\psi_{(n_2,l_2,m_2)}\;,\label{state0}\\
E&=&E_{n_1}+E_{n_2}\;=\;-\frac{Z^2}{2}\frac{me^4}{\hbar}\left[\frac{1}{n_1}+\frac{1}{n_2}\right]=-{Z^2}E_{\rm H}
\left[\frac{1}{n_1}+\frac{1}{n_2}\right]\;,\label{energy0}
\end{eqnarray}
where
\begin{eqnarray}
\left[\frac{\bar{p}_i^2}{2m_i}-\frac{e}{r_i}\right]\psi_{(n_i,l_i,m_i)} = E_{n_i}\psi_{(n_i,l_i,m_i)}\;\;;\;\;i=1,2\;,
\end{eqnarray}
and where $n$, $l$, $m$ denote main, orbital and azimuthal quantum number respectively. However,
in order to find the whole spectrum of the model, i.e., in order to solve the following equation
\begin{eqnarray}
H\psi = \left(H_0+H_{\rm int}\right)\psi_{\rm tot}  = E_{\rm tot}\psi_{\rm tot}
\end{eqnarray}
with $H_{\rm int}\ne 0$, one can use the perturbation expansion method \cite{kiku}. Then, the first-order corrections to the energy levels (\ref{energy0}) look as follows
\begin{eqnarray}
E^{(1)} &=& \left(\psi,H_{\rm int}\psi\right)\;,\label{expectation}
\end{eqnarray}
while
\begin{eqnarray}
\psi_{\rm tot} &=& \psi + \lambda\psi^{(1)} +  \lambda^2\psi^{(2)} + \cdots\;,\\
E_{\rm tot} &=& E + \lambda E^{(1)} + \lambda^2 E^{(2)} + \cdots\;.
\end{eqnarray}
Particularly, for ground state energy described by the set $n=1$, $l=m=0$ we have\footnote{$a_0 = 0.5{\AA}$ denotes the Bohr radius.}
\begin{eqnarray}
\psi(r_1,r_2) = \psi_{100}(r_1)\psi_{100}(r_2) = \frac{1}{\pi}\left[\frac{Z}{a_0}\right]^3{\rm e}^{-Z{(r_1+r_2)}/{a_0}}\;,\label{ground}
\end{eqnarray}
and, by simple integration we obtain
\begin{eqnarray}
E^{(1)} = E_{100}^{(1)} = \frac{5Z}{4}E_{\rm H}\label{correction}
\end{eqnarray}
Consequently, the total energy of the system is given by
\begin{eqnarray}
E_{\rm ground} = -2{Z^2}E_{\rm H}+\frac{5Z}{4}E_{\rm H}\;,\label{total}
\end{eqnarray}
while in the case of Helium atom $(Z=2)$, it is equal to\footnote{$E_{\rm H}=\frac{e^2}{2a_0}=13.605$ eV.}
\begin{eqnarray}
E_{\rm Helium} = -8E_{\rm H}+\frac{10}{4}E_{\rm H} = -108.84\; eV +34.01\; eV = -74.83 \;eV\;.\label{totalhelium}
\end{eqnarray}
However, unfortunately, it should be also noted, that the above result is different than the experimental one, in accordance with which
\begin{eqnarray}
E_{\rm experimental} = -79.03 \;eV\;,\label{experimental}
\end{eqnarray}
and finally, we get\footnote{The more precise outcomes were obtained in the framework of so-called variational method \cite{lit1}, where $E_{\rm Helium} = -78.95\; eV$ and $\Delta E = 0.08\; eV$. Hence, it seems quite sensible to analyze the noncommutative correction using variational approach. However, the main aim of this article concerns the exploration of perturbation scheme, and just for this reason, the studies in variational groundwork are omitted  and postponed for further investigation.}
\begin{eqnarray}
\Delta E = E_{\rm Helium} - E_{\rm experimental} = -74.83\; eV + 79.03\; eV = 4.20 \;eV\;.\label{finally}
\end{eqnarray}

\section{Two-particle $(M=2)$ Coulomb system in twisted space-time}

Let us now turn to the main aim of our investigations, i.e., to the construction of two-particle Coulomb system for quantum space-times (\ref{spaces300}).
In first step of our construction, we extend the twisted space to the whole algebra of momentum and position operators as follows
\begin{eqnarray}
&&[\;\hat{ x}_{1A},\hat{ x}_{2B}\;] = if_{\kappa_a}({t})\delta_{AB}\;\;\;,\;\;\;
[\;\hat{ p}_{iA},\hat{ p}_{jB}\;] = 0 = [\;\hat{ x}_{iA},\hat{ x}_{3B}\;]\;,\label{rel1}\\
&&\nonumber\\
&&~~~~~~~~~~~~~[\;\hat{ x}_{iA},\hat{ p}_{jB}\;] =
i\hbar\delta_{ij}\delta_{AB}\;.\nonumber
\end{eqnarray}
One can check that the above relations satisfy the Jacobi identity and for deformation parameter
$\kappa_a$ approaching zero become classical. \\
Next, we define the proper Hamiltonian operator in a standard way by\footnote{We define the Hamiltonian operator by replacement in the formula (\ref{ham}) the classical operators $(x_i,p_i)$ by their noncommutative counterparts $(\hat{x}_i,\hat{p}_i)$.}
\begin{eqnarray}
\hat{H} = \frac{1}{2m_{\rm e}}\left[\;{\bar{\hat{p}}_1^2}+{\bar{\hat{p}}_2^2}\;\right]-\frac{Ze^2}{\hat{r}_1}-
\frac{Ze^2}{\hat{r}_2} + \frac{e^2}{\hat{r}_{12}}\;,\label{twistzdirection}
\end{eqnarray}
and in order to perform the basic analyze of the system, we represent the
noncommutative variables $({\hat x}_i, {\hat p}_i)$ by the classical
ones $({ x}_i, { p}_i)$ as  (see e.g.
\cite{romero1}-\cite{kijanka})
\begin{eqnarray}
{\hat x}_{1A} &=& { x}_{1A} - {{f_{\kappa_a}(t)}}/{(4\hbar)}p_{2A}\;\label{rep1}\\
{\hat x}_{2A} &=& { x}_{2A} + {{f_{\kappa_a}(t)}}/{(4\hbar)}p_{1A}
\;,\label{rep2}\\
{\hat x}_{3A} &=& { x}_{3A}\;\;\;,\;\;\;{\hat p}_{iA} \;=\; { p}_{iA} \;,
\label{rep3}
\end{eqnarray}
where
\begin{equation}
[\;x_{iA},x_{jB}\;] = 0 =[\;p_{iA},p_{jB}\;]\;\;\;,\;\;\; [\;x_{iA},p_{jB}\;]
={i\hbar}\delta_{ij}\delta_{AB}\;. \label{classpoisson}
\end{equation}
Then, the  Hamiltonian (\ref{twistzdirection}) takes the form
\begin{eqnarray}
\hat{H}(t) &=& \frac{1}{2m_{\rm e}}\left[\;{\bar{{p}}_1^2}+{\bar{{p}}_2^2}\;\right]\;-\;{Ze^2}\sum_{A=1}^2\left[{{({ x}_{1A} - {{f_{\kappa_a}(t)}}/{(4\hbar)}p_{2A})^2}}\;+\right.\nonumber\\
&+&
\left.({ x}_{2A} + {{f_{\kappa_a}(t)}}/{(4\hbar)}p_{1A})^2+x_{3A}^2\right]^{-\frac{1}{2}}\;+\nonumber\\
&+&e^2\left[({ x}_{11} - {{f_{\kappa_a}(t)}}/{(4\hbar)}p_{21} - { x}_{12} + {{f_{\kappa_a}(t)}}/{(4\hbar)}p_{22})^2\right.+\label{newform}\\
&+&({ x}_{21} + {{f_{\kappa_a}(t)}}/{(4\hbar)}p_{11} - { x}_{22} - {{f_{\kappa_a}(t)}}/{(4\hbar)}p_{12})^2+\nonumber\\
&+&\left.(x_{31}-x_{32})^2\right]^{-\frac{1}{2}}\;,\nonumber
\end{eqnarray}
while for small values of deformation function ${{f_{\kappa_a}(t)}}$, it looks as follows
\begin{eqnarray}
\hat{H}(t) &=& H_0 + \hat{H}_{\rm int}(t) = \sum_{A=1}^2\frac{\bar{p}^2_A}{2m_{\rm e}} - \sum_{A=1}^2\frac{Ze^2}{r_A} + \frac{e^2}{r_{12}} - \sum_{A=1}^2\frac{Ze^2}{4\hbar}\frac{f_{\kappa_a}(t)}{r_A^3}L_{3A}\;+\label{joachim}\\
&+& \frac{e^2f_{\kappa_a}(t)}{4\hbar r_{12}^3}\left[\;L_{31}+L_{32}\;\right] - \frac{e^2}{4\hbar r_{12}^3}\left[\;G_{12}+G_{21}\;\right] + {\cal O}(f_{\kappa_a}^2(t))\;,\nonumber
\end{eqnarray}
with $L_{3A} = x_{1A}p_{2A}-x_{2A}p_{1A}$, $G_{AB} = x_{1B}p_{2A}-x_{2B}p_{1A}$, and with symbol ${\cal O}(f_{\kappa_a}^2(t))$ denoting the elements at least quadratic in parameter $\kappa_a$. Besides, it should added, that present
in the above formula term $\hat{H}_{\rm int}(t)$ describes the new, deformed interaction vertex, which defines the total spectrum of the model by
\begin{eqnarray}
\hat{H}(t)\psi(t) = E(t)\psi(t)\;.\label{joachim1}
\end{eqnarray}
Next, in order to solve the equation (\ref{joachim}) with respect the energy $E(t)$, we assume that
\begin{eqnarray}
\psi(t) &=& \psi + \lambda\psi^{(1)}(t) + \lambda^2\psi^{(2)}(t)+ \cdots\;,\\
E(t) &=& E + \lambda E^{(1)}(t) + \lambda^2E^{(2)}(t)+ \cdots\;,
\end{eqnarray}
where $E$ and $\psi$ are both given by the formula (\ref{dixiland}); then, we have
\begin{eqnarray}
E^{(1)}(t) &=& (\psi,\hat{H}_{\rm int}(t)\psi)\;.
\end{eqnarray}
Further, using the result (\ref{correction})
we get
\begin{eqnarray}
\hat{E}^{(1)}(t) = \hat{E}_{100}^{(1)}(t) = \frac{5Z}{4}E_{\rm H} + \frac{e^2f_{\kappa_a}(t)}{4\hbar}\sum_{i=1}^4 I_i\;,\label{thetacor}
\end{eqnarray}
what leads to the total $f_{\kappa_a}(t)$-deformed ground energy of the system of the form\footnote{${\rm d}\Gamma \equiv \prod_{A=1}^2\cos\theta_A d\theta_A d\varphi_A r_A^2dr_A.$}
\begin{eqnarray}
&&E_{{\rm ground},\kappa_a} = -2{Z^2}E_{\rm H}+\frac{5Z}{4}E_{\rm H}+ \frac{e^2f_{\kappa_a}(t)}{4\hbar}\sum_{i=1}^4 I_i\;;\label{en}\\
&&I_{A}=-Z\int {\rm d}\Gamma\;\psi^{*}[{L_{3A}}/{r_A^3}]\psi\;\;;\;\;A=1,2\;,\\
&&I_3=\int {\rm d}\Gamma\;\psi^{*}[{(L_{31}+L_{32})}/{r_{12}^3}]\psi\;,\\
&&I_4=-\int {\rm d}\Gamma\;\psi^{*}[{(G_{12}+G_{21})}/{r_{12}^3}]\psi\;.
\end{eqnarray}
It should be noted, however, that three first integrals $I_1$, $I_2$ and $I_3$ vanish since $L_{3A}\Psi(r_1,r_2)$ $= 0$. Similarly, by direct nontrivial calculation one can also check, that the last one $(I_4)$ is equal to zero as well\footnote{$p_{iA}\Psi(r_1,r_2)=-i\hbar\frac{\partial}{\partial x_{iA}}\Psi(r_1,r_2)$.},\footnote{The vanishing of $I_4$-term is with accordance with fact, that since tensors $G_{12}$ and $G_{21}$ are linear in momentum operator, the last integral remains complex.}. Hence, we have
\begin{eqnarray}
E_{{\rm corrections}, \kappa_a} = \frac{e^2f_{\kappa_a}(t)}{4\hbar}\sum_{i=1}^4 I_i = 0\;,
\end{eqnarray}
and, consequently
\begin{eqnarray}
E_{{\rm Helium},\kappa_a} = E_{{\rm Helium}} = -74.83\;eV\;.\label{encomp}
\end{eqnarray}
Obviously, for function ${{f_{\kappa_a}(t)}}=0$ the above formulas become undeformed.

\section{Final remarks}

In this article we provide two-particle system in Coulomb potential defined on twist-deformed space-time (\ref{spaces300}) with $A,B=1,2$. Particularly, we demonstrate, that for small values of deformation function ${{f_{\kappa_a}(t)}}$, the obtained first-ordered corrections for ground Helium energy vanish. In such a way we show, that the impact of noncommutativity appears just at the second order of the quantum-mechanical perturbation. It should be noted, however, that such an expectation requires the additional investigation, which seems to be very difficult from technical point of view, and for this reason, it is postponed for further publication\footnote{In the case of the second-order corrections, there appear the objects of the following form $$(\psi_{(n,l,m)}\psi_{(n',l',m')},\hat{H}_{\rm int}(t)\psi_{(1,0,0)}\psi_{(1,0,0)})$$ where $\psi_{(n,l,m)}\psi_{(n',l',m')}$ are different than ground state. Unfortunately, their calculation in our case seems to be quite complicated from technical point of view, and for this reason, it is postponed for further investigation.}.

\section*{Acknowledgments}

The author would like to thank J. Lukierski for valuable discussions.

\end{document}